\documentclass[%
 reprint,
 amsmath,amssymb,
 aps,
prb,
]{revtex4-2}
\usepackage{float}
\usepackage{svg}

\usepackage[dvipsnames]{xcolor} 

\definecolor{myblue}{RGB}{0,80,190}  

\usepackage[colorlinks=true,
            linkcolor=myblue,
            citecolor=myblue,
            urlcolor=myblue,
            pdfusetitle]{hyperref}

\usepackage{graphicx}
\usepackage{dcolumn}
\usepackage{bm}
\usepackage[dvipsnames]{xcolor}
\usepackage{braket}

\newcommand{\rmi}{\text{i}}

\begin{document}
\preprint{APS/123-QED}

\title{Emergence of Turbulence in a counterflow geometry of 2D Polariton Quantum Fluids}

\author{L. Depaepe}
\affiliation{Univ. de Lille, CNRS, UMR 8523–PhLAM–Physique des Lasers, Atomes et Molécules, Lille, France\\}%
\author{K. Ouahrouche}
\affiliation{Univ. de Lille, CNRS, UMR 8523–PhLAM–Physique des Lasers, Atomes et Molécules, Lille, France\\}%
\author{A.~Amo}
\affiliation{Univ. de Lille, CNRS, UMR 8523–PhLAM–Physique des Lasers, Atomes et Molécules, Lille, France\\}%

\author{C. Hainaut}
\email{clement.hainaut@univ-lille.fr}
\affiliation{Univ. de Lille, CNRS, UMR 8523–PhLAM–Physique des Lasers, Atomes et Molécules, Lille, France\\}%

\date{\today}
\begin{abstract}
We numerically investigate the nonlinear dynamics of a two-dimensional exciton–polariton quantum fluid coherently driven by two counter-propagating laser beams. Using an exciton–photon coupled driven–dissipative Gross–Pitaevskii framework, we identify four distinct regimes—linear, solitonic, turbulent, and superfluid-emerging from the interplay between pump strength, laser detuning, and injected momentum, which together control the balance between kinetic and interaction energies in the quantum fluid. The different regimes are characterized through real-space and momentum-space observables, as well as through the temporal first-order coherence function. We show that turbulence occupies a well-defined and extended region of parameter space, marked by spontaneous vortex nucleation,  and a pronounced reduction of temporal coherence, providing a clear signature of nonstationary dynamics. By constructing quantitative phase diagrams, we delineate the transitions between the various regimes and identify multiple pathways connecting solitonic, turbulent, and superfluid behaviors. Finally, we demonstrate that the turbulent regime persists over experimentally realistic parameter ranges compatible with state-of-the-art GaAs-based micro-cavity platforms, establishing counter-propagating polariton flows as a robust and versatile setting for the study of driven–dissipative quantum turbulence in two dimensions.

\end{abstract}

\maketitle


\section{Introduction}
Quantum fluids, in which macroscopic coherence emerges from the collective behavior of bosonic particles, display a wide range of dynamical phenomena, including superfluidity, soliton formation, and quantum turbulence. These effects were first identified through the existence of quantized vortices in superfluid helium \cite{Hall1956,Leggett2004}. More recently, the experimental realization of Bose–Einstein condensates (BECs) \cite{Anderson1995,Davis1995} has established ultracold atomic gases \cite{Giorgini2008,Bloch2008} as a powerful platform for studying superfluidity and quantum turbulence.

Quantum turbulence is characterized by disordered flow patterns, spontaneous quantized vortex nucleation, and complex spatiotemporal dynamics that enable energy transfer across scales. In classical turbulence, Richardson’s direct energy cascade \cite{Richardson2007}, where energy flows from large to small length scales, is a well-established phenomenon observed across many physical systems. A similar direct cascade has also been reported in quantum systems, notably in ultracold atomic gases \cite{Navon2019}. In two dimensions, however, turbulence exhibits a distinctive feature: the possibility of an inverse energy cascade \cite{Kraichnan1967,Reeves2013}, in which energy is transferred from small to large scales. This process enables vortex clustering \cite{Bradley2012,Gauthier2019,Johnstone2019} via a phenomenon known as Onsager vortex formation  \cite{Onsager1949}. Recently, inverse energy cascades have been experimentally observed in driven homogeneous Bose gases \cite{Karailiev2024}.

In parallel, quantum fluids of light \cite{Carusotto2013}, based on exciton-polariton bosonic quasiparticles arising from the strong coupling between cavity photons and excitons confined in semiconductor quantum wells \cite{Weisbuch1992}, have emerged as a promising platform to explore quantum fluid physics in two-dimensional driven-dissipative systems. In these systems, steady states result from a balance between driving and dissipation, making polariton fluids particularly suitable for investigating properties of non-equilibrium condensates \cite{Fontaine2022,Amo2009,Wertz2010,Stepanov2019,Caputo2017} and out-of-equilibrium quantum hydrodynamics.

As in conservative quantum fluids, frictionless flow has been experimentally demonstrated in nonlinear optical systems \cite{Wan2006,Bolda2001} and exciton-polariton systems \cite{Carusotto2004,Amo2009,Amo2009Nat,Lerario2017,Peng2022}, signaling superfluid behavior. Subsequent studies have explored various hydrodynamic phenomena, including quantized vortices \cite{Lagoudakis2008,Nardin2011,Sanvitto2010}, vortex dynamics \cite{Lagoudakis2011,Hivet2012,Simula2014}, and soliton formation and stability \cite{Amo2011,Sich2011,Matre2020,Koniakhin2019,Snake_instability}. Together, these pioneering works have established a solid understanding of out-of-equilibrium quantum hydrodynamics. 

The onset of turbulent flows in polariton systems has primarily been studied in obstacle-flow geometries \cite{Amo2009,Amo2011,Lerario2017,Peng2022,Simu_Pigeon,Disorder_Amelio}, where a polariton fluid interacts with a localized defect and gradually decays during propagation. However, a systematic investigation of turbulence in more symmetric and confined configurations remains scarce \cite{Panico2023}. Such geometries would provide access to energy cascades across multiple length scales, as well as to microscopic vortex dynamics and clustering phenomena. In this context, counter-propagating laser beams that inject two coherent polariton fluids constitute a particularly promising configuration, offering a tunable, extended, high-density interaction region with independent control over the injected momentum and energy.
 
In this work, we numerically investigate the emergence of turbulence and coherent structures in a two-dimensional polariton fluid driven by counter-propagating coherent pumps. Using an exciton–photon coupled driven-dissipative Gross–Pitaevskii framework, we simulate the system dynamics over a broad range of parameters and identify four distinct dynamical regimes: linear, solitonic, turbulent, and superfluid. These regimes are characterized using real-space and momentum-space observables, as well as statistical indicators such as the temporal first-order coherence function. Our analysis shows that the different regimes originate from the competition between kinetic and interaction energies within the quantum fluid. Furthermore, we demonstrate that the transitions between regimes are governed by the interplay of nonlinearity, pump strength, and laser detuning, allowing us to construct quantitative phase diagrams that describe distinct domains of existence. Finally, we demonstrate that an extended turbulent regime exists within experimentally realistic parameter ranges that are accessible in state-of-the-art polariton platforms. 

The paper is organized as follows. We first introduce the theoretical model, the conventions adopted throughout the work, and the counter-propagating pumping geometry. We then present the first results of long-time numerical simulations, enabling a clear identification of the four dynamical regimes. Next, we explain how we construct a phase diagram and discuss the various possible transitions.

\section{Model and Geometry}

\subsection{Model equation}

\begin{figure}[t!]
    \centering
    \includegraphics[width=1\linewidth]{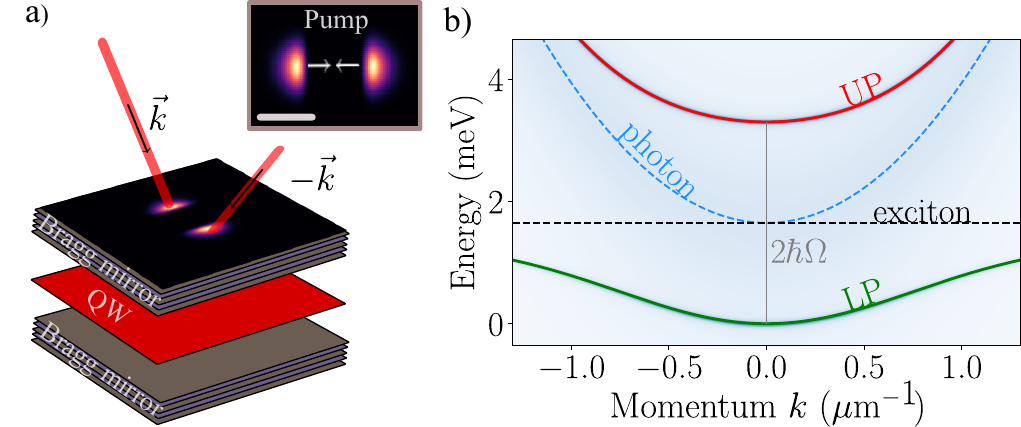}
    \caption{(a)  Schematic of the experimental two-dimensional polariton cavity, showing two counter-propagating pump beams and a zoom of the 2D pump intensity profile. The scale bar corresponds to approximately $54\,l_0' \approx 75~\mu\mathrm{m}$. (b) Illustration of the bare exciton and cavity mode dispersions (dashed lines), together with the lower (LP) and upper (UP) polariton dispersions (solid lines). The depicted dispersions correspond to a system with a Rabi splitting of
    $2\hbar\Omega' = 3.3~\mathrm{meV}$ and an initial $50{:}50$ exciton--photon
    composition, which occurs when the $k=0$ photon energy equals the exciton
    energy.}
    \label{fig:Pump+cavity_scheme}
\end{figure}

Exciton-polaritons are bosonic quasiparticles resulting from the strong coupling between cavity photons and excitons, as schematically illustrated in the left panel of Fig.\hyperref[fig:Pump+cavity_scheme]{~\ref*{fig:Pump+cavity_scheme}a}. In a typical experiment setup, optical confinement is achieved using two Distributed Bragg Reflectors (DBRs), which trap photons at the microscale and endow them with an effective mass. In addition, one or several quantum wells hosting excitonic excitations are placed at the antinodes of the cavity electromagnetic field, allowing the system to reach the strong-coupling regime. Owing to their photonic component, polaritons exhibit picosecond-scale dynamics and can be optically manipulated and probed with high spatial and temporal resolution. Their excitonic component provides nonlinear interactions, enabling the emergence of collective phenomena such as superfluidity and soliton formation.

In the nonlinear driven–dissipative regime, the dynamics of the excitonic field $\psi_X(\mathbf r,t)$ and the cavity photon field $\psi_C(\mathbf r,t)$ are accurately described by treating them as strongly coupled fields. This approach is more general than the generalized Gross–Pitaevskii equation, which restricts the description to the lower polariton branch and is valid only when the Rabi coupling energy largely exceeds all other relevant energy scales \cite{RevModPhys.85.299}. The coupled exciton–photon framework incorporates additional physical effects, namely the saturation of exciton–exciton interactions and the momentum dependence of the polariton composition (Hopfield coefficients), both of which can significantly modify the system’s dynamical behavior.

Within the mean-field approximation, the system dynamics are therefore governed by a set of coupled equations written in the rotating frame of the pump field. For clarity, we use their normalized (dimensionless) form, given in Eq.~\ref{eq:CoupledEqADim}. The derivation of these equations and their connection to the dimensional formulation are detailed in Appendix~\ref{app:adim}. Throughout the manuscript, quantities carrying a prime ($'$) denote dimensional variables and parameters, while unprimed quantities are dimensionless.

\begin{equation}
\label{eq:CoupledEqADim}
\begin{aligned}
\rmi\frac{\partial\psi_C}{\partial t} 
= &\left(-\Delta_C - \nabla^2 + W(\mathbf{r})\right)\psi_C 
- \rmi \frac{\gamma_C}{2} \psi_C 
\\&+ \rmi \sqrt{\frac{\gamma_C}{2}} F_{\text{inc}}(\mathbf{r}) 
- 2\psi_X,\\
\rmi\frac{\partial\psi_X}{\partial t} 
=& \left(-\Delta_X + |\psi_X|^2\right)\psi_X 
- \rmi \frac{\gamma_X}{2} \psi_X 
- 2\psi_C.
\end{aligned}
\end{equation}
Here, \( \gamma_{C,X} \) denote the decay rates of cavity photons and excitons respectively; \( \Delta_{C,X} = \omega_{C,X} - \omega_p \) are the corresponding detunings with respect to the pump angular frequency \( \omega_{\text{p}} \). The function \( F_{\mathrm{inc}}(\mathbf{r}) \) describes the spatial amplitude profile of the incident coherent field, and $W(\mathbf{r})$ accounts for spatial disorder. In Eq.~\ref{eq:CoupledEqADim}, the Laplacian operator 
$\nabla^2$ appears only in the photonic equation, since the exciton mass is sufficiently large that its dispersion can be considered flat over the range of transverse momenta relevant to the present problem (see Fig.\hyperref[fig:Pump+cavity_scheme]{~\ref*{fig:Pump+cavity_scheme}b}). For simplicity, we restrict ourselves to equal decay rates, \( \gamma_C = \gamma_X \) , and include the cavity–pump coupling factor \( \sqrt{ \gamma_C/ 2} \) multiplying the driving term, following the standard input–output formalism \cite{Millburn}.

In this work, we further focus on the resonant case in which the bare exciton and photon energies at in-plane wavector $k=0$ coincide: $\hbar\omega'_X(0)=\hbar\omega'_C(0,k=0)=\hbar\Omega'$ (see Fig.\hyperref[fig:Pump+cavity_scheme]{~\ref*{fig:Pump+cavity_scheme}b}). This condition corresponds to a 50:50 exciton–photon mixing at zero momentum. In the absence of losses, driving field and nonlinearities, the resulting lower (LP) and upper (UP) polariton dispersion relations \cite{Carusotto2013} are given by

\begin{align}
E'_{\substack{\mathrm{LP} , \mathrm{UP}}}(k)= 
&\frac{1}{2}\Big[ \hbar\omega'_C(k) +\hbar \omega'_X \notag\\
&\pm \sqrt{ \big( \hbar\omega'_C(k) - \hbar\omega'_X \big)^2 + 4\hbar^{2}\Omega'^{2}} \Big],
\label{eq:L_dispersion}
\end{align}

\noindent where the momentum dependence of the excitonic dispersion has been neglected due to the large exciton mass. The corresponding dispersion curves are shown in Fig.\hyperref[fig:Pump+cavity_scheme]{~\ref*{fig:Pump+cavity_scheme}b}, with the energy reference chosen at the lower polariton energy $E'_{\substack{\mathrm{LP}}}(k=0)$

To retain broad applicability while ensuring experimental relevance and provide a typical situation allowing visualization of the problem, we provide, throughout the paper, both dimensionless parameters and their physical counterparts. The Rabi splitting is fixed by $2\hbar \Omega' = 3.3~\text{meV}$, which sets the characteristic timescale $\tau'_0 = 2/\Omega' \approx 0.8~\text{ps}$ as well as a typical energy scale $\tilde{E}'_0=\hbar\Omega'/2=0.83$ meV. The cavity loss rate is $\gamma_c' = (40~\text{ps})^{-1}$,  and the exciton–exciton interaction strength is $\hbar g_X' = 0.003~\text{meV}\,\mu\text{m}^2$. These parameters define a characteristic length scale $l_0' = \sqrt{\hbar/(m'_C\Omega')}= 1.38~\mu\text{m}$
where the photon effective mass is chosen as 
$m'_C=2.4\times10^{-5}m'_e$.

The term \( W(\mathbf{r}) \) in Eq. \ref{eq:CoupledEqADim} represents a spatially disordered potential, known to arise in polariton experiments \cite{Houdr2000,Gurioli2005,Amo2009disorder} due to slight variation of the index of refraction in microcavities due to unavoidable imperfections in cavity fabrication processes. The latter acts exclusively on the photonic field. In this work, following \cite{Disorder_Amelio}, we model the structural disorder as a Gaussian random potential \( W(\mathbf r) \) with zero mean, \( \langle\!\langle W(\mathbf r) \rangle\!\rangle = 0 \) and with spatial correlations given by
\begin{equation}
\langle\!\langle W(\mathbf r_1) W(\mathbf r_2) \rangle\!\rangle = W_0^2 \exp\!\left( -\frac{|\mathbf r_1 - \mathbf r_2|^2}{4\sigma_W^2} \right),
\label{eq:disorder}
\end{equation}
where \( \langle\!\langle \cdot \rangle\!\rangle \) denotes an average over disorder realizations. As in  \cite{Disorder_Amelio} we chose a correlation length $\sigma_W = 0.36$ (\( \sigma_W' = 0.5\,\mu\mathrm{m}\)) and a relative amplitude $W_0 = 1.43 \times 10^{-3} $ (\( \hbar W_0' = 1.2\,\mu\mathrm{eV}\)) which remains much smaller than all other relevant energy scales and therefore acts as a weak perturbation.

The system dynamics are obtained by numerically integrating Eq.~\ref{eq:CoupledEqADim}. As discussed in Ref.~\cite{Disorder_Amelio}, the presence of disorder breaks the spatial symmetries of the system and constitutes an essential ingredient for triggering dynamical instabilities and turbulent behavior \cite{Disorder_Amelio,exp_VortexFromSoliton_counterflow}. It is important to note that, in nonlinear simulations, unavoidable numerical errors may also act as effective symmetry-breaking seeds and artificially induce turbulent-like dynamics, as it was previously the case in Ref.~\cite{Simu_Pigeon}. In the present work, we employ a disorder strength compatible with experimental values and verify numerically that the emergence of turbulence is independent of the temporal and spatial discretizations as well as of the grid size. This ensures that the observed turbulent behavior originates from physical disorder rather than numerical artifacts.

\subsection{Counterpropagating pump geometry}

We employ a counter-propagating injection scheme in which two coherent laser beams drive the polariton fluid with equal amplitudes and opposite in-plane momenta, as illustrated in Fig.\hyperref[fig:Pump+cavity_scheme]{~\ref*{fig:Pump+cavity_scheme}a}. The pump profiles are constructed by truncating Gaussian beams with a sharp spatial cutoff. The exact mathematical expression of the pump profiles used in the numerical simulations is provided in Appendix~\ref{app:pump_profile}. 
Such steep-edged pump profiles are commonly employed in experiments as a compromise between minimizing phase imprinting from the external drive and maintaining sufficient optical injection into the cavity. This approach allows the polariton fluid to evolve according to its intrinsic dynamics, while remaining compatible with experimental power constraints \cite{Simu_Pigeon,pump_profile1}.
The two lobes are separated by a peak-to-peak distance of $d' = 57\,l_0' \approx 80~\mu\mathrm{m}$, leaving an empty central region of width approximately $25\,l_0' \approx 35~\mu\mathrm{m}$. This separation is chosen to balance two competing requirements: it provides ample space for the nonlinear dynamics to develop in the central interaction region, while remaining compatible with the finite polariton lifetime, ensuring that the two counter-propagating flows reach the overlap region without undergoing significant decay.

\section{Results: Identification of four distinct Regimes}

The dynamics of the coupled dimensionless equations~\ref{eq:CoupledEqADim} in a counter-propagating configuration are computed using a Fourier split-step method (Code available in open access GitHub repository \cite{Depaepe_Code}). To suppress nonphysical reflections and recirculation at the boundaries, which is a standard concern in 2D driven-dissipative simulations, spatially varying losses are applied near the edges. Zero-padding in Fourier space is also used to minimize aliasing and other frequency artifacts.

Simulations are performed on a grid of $N \times N$ points with $N=512$, spanning a square of length $L \times L$ with $L=256$ ($L'=256\,l_0'$), giving a pixel size $\Delta x=1/2$ ($\Delta x' = l_0'/2$). This choice ensures that the relevant spatial features of the polariton fluids are well resolved while keeping the computational cost reasonable. \\
Starting from an empty system, the pump is switched on, up to a constant value, using a smooth exponential ramp $f(t)=1-\exp(-t/\tau)$ with $\tau=70\,\tau_0'$, corresponding to an effective ramp duration $\sim 3\tau \simeq 210\,\tau_0'$ (reaching $95\%$ of the final value), which largely exceeds the relaxation times $1/\gamma_{C,X}=50\,\tau_0'$. The system is evolved up to a total time
$t' \gtrsim 2000\,\tau_0'$, including the ramp, which is sufficient to reach stationary states whenever they exist. Convergence with respect to the system size, spatial resolution, and time step has been carefully checked to ensure the robustness of the observed dynamics (see Appendix~\ref{app:code_verif}).

\begin{widetext}
\centering

\begin{figure}[H]
    \includegraphics[width=\textwidth]{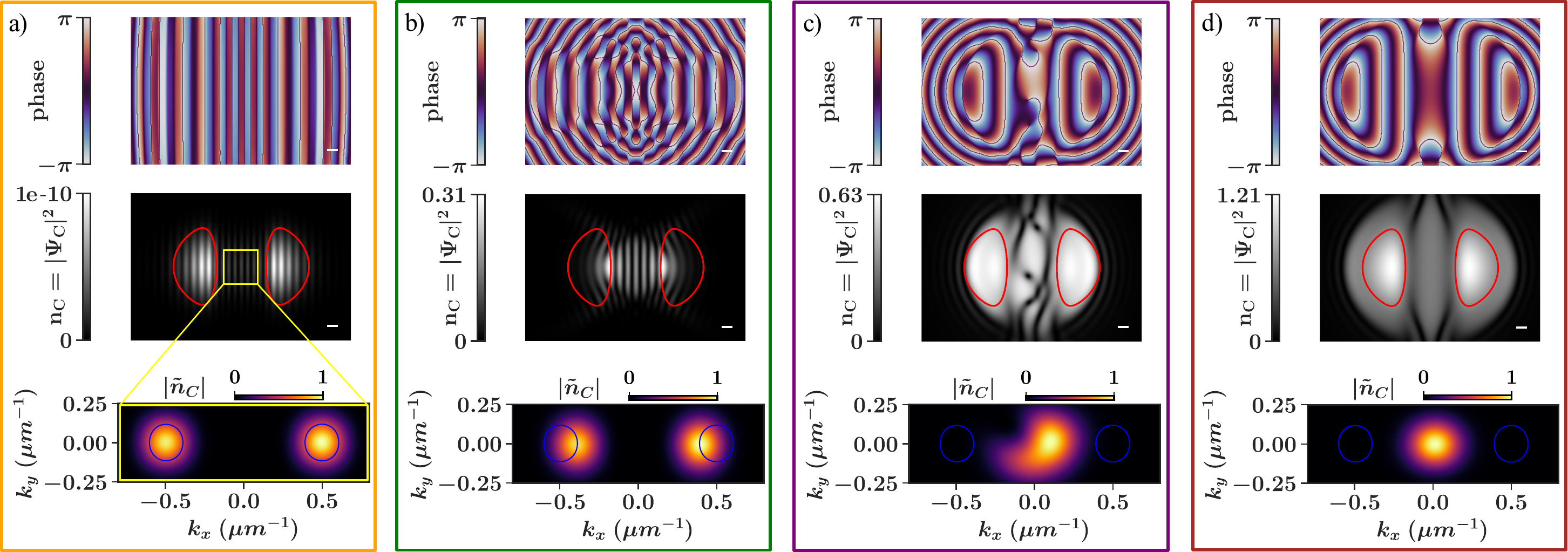}
    \caption{Four distinct regimes are shown: (a) linear ($F_{\text{inc}} = 10^{-3}$), (b) solitonic ($F_{\text{inc}} = 0.6$), (c) turbulent ($F_{\text{inc}} = 1.2$), and (d) superfluid ($F_{\text{inc}} = 3.7$).
    For each regime, the top panels display the two-dimensional phase distribution of the photonic field. The middle panels show the two-dimensional photonic density $n_C = |\Psi_C|^2$, with the pump positions indicated in red at the $1/e^2$ contour of its maximum intensity. The bottom panels present $\tilde n_C$, the normalized momentum-space distribution of the photonic field in the central region (highlighted by the yellow rectangle in panel a)), obtained from the two-dimensional Fourier transform of the real-space photonic density, $n_C$.
    The scale bar corresponds to $10\,\mu\mathrm{m}$. 
    Parameters: Rabi splitting $2\hbar\Omega' = 3.3$~meV, pump momenta 
    $k_p = 0.4 \leftrightarrow k_p' = 0.29~\mu\mathrm{m}^{-1}$, laser detuning $\Delta=0.22\leftrightarrow\Delta'=0.18\,\mathrm{meV}$, and loss rates 
    $\gamma = 0.02 \leftrightarrow \gamma' = 0.025~\mathrm{ps}^{-1}$.}
    \label{fig:4regimes}
\end{figure}
\end{widetext}

Figure~\ref{fig:4regimes} shows a snapshot in the long-time behavior of the photonic densities (middle panels), phase profiles (upper panels), and momentum spectra (lower panels). The momentum spectra are obtained by Fourier transforming the real-space photonic density in the central region of the fluid, indicated by the yellow square, which defines our region of interest. In all four examples (from (a) to (d)), the detuning of the external laser with respect to the lower polariton branch, defined as $\Delta'=E'_{\text{L}}-E'_{\text{LP}}(k=0)$, is fixed at $\Delta' = 0.22 \tilde E_0' \approx 0.18~\mathrm{meV}$ with $E'_{\text{L}}$ the energy of the pump laser drive. With this definition, a positive detuning means that the pump energy is higher than the energy of the lower polariton. In addition, for the four panels, the pump wavevector is chosen at $k_{\text{p}}' = 0.4\,l_0'^{-1} \approx 0.29~\mu\mathrm{m}^{-1}$ and the pump amplitude $F_{\text{inc}}$ is varied from $10^{-3}$ to 6. The red delimitation in the real-pace panels is an indicate for the eyes of the region of the external pump driving satisfying the conditions of Fig. \ref{fig:Pump+cavity_scheme}a).

At low pump amplitude, $F_{\text{inc}} = 10^{-3}$ (Fig.\hyperref[fig:4regimes]{~\ref*{fig:4regimes}a}), the system forms a regular standing-wave pattern with alternating $\pi$ phase shifts between adjacent fringes, which is the hallmark of the linear regime. The characteristic wavevector $k_{\mathrm{linear}}'=0.7\,l_0'^{-1}\approx0.5\mathrm{\mu m^{-1}}$ associated with this standing wave is determined by the polariton dispersion relation given in Eq.~\ref{eq:L_dispersion}. 


Increasing the pump amplitude to $F_{\text{inc}} = 0.6$  (Fig.\hyperref[fig:4regimes]{~\ref*{fig:4regimes}b}) enhances the fluid density in the central region. As a result, nonlinear interactions set in, which result in a distortion of the interference pattern, as clearly visible in the phase distribution. Despite these distortions, the system retains a robust density-modulated structure, which we identify as the solitonic regime. Compared to the linear case, the fringes in this regime are more widely spaced, indicating a reduction of the characteristic wavevector, $k_{\mathrm{solitonic}} < k_{\mathrm{linear}}$. This shift is also apparent in momentum space, where the blue circle marks the wavevector distribution associated with the linear regime for reference.

At a higher pump value, \(F_{\text{inc}} = 1.2\) (Fig.\hyperref[fig:4regimes]{~\ref*{fig:4regimes}c}), the system enters a turbulent regime where no stationary state can be reached. This regime is characterized by spontaneous vortex formation and complex spatiotemporal dynamics, accompanied by strong spatiotemporal fluctuations in both phase and density. 
The onset of turbulence can be traced back to the dynamical instability of solitonic structures, known as the snake instability \cite{Snake_instability}, whereby solitons destabilize and decay into multiple vortex–antivortex pairs. In the counter-propagating geometry considered here, soliton-like structures forming near the pump regions repeatedly undergo this fragmentation process, continuously generating vortices that propagate and interact within the central overlap region. A representative sequence of snapshots illustrating this soliton breakup is provided in the Appendix E. The density distribution in Fig.\hyperref[fig:4regimes]{~\ref*{fig:4regimes}c} is asymmetric due to the presence of structural disorder (Eq.~\ref{eq:disorder}), which breaks the parity symmetry of the underlying equations. In addition, vortex nucleation and strong nonlinear interactions lead to a redistribution of energy across momentum space, resulting in the broader spectral distribution centered around \(\mathbf{k}=0\) observed in Fig.\hyperref[fig:4regimes]{~\ref*{fig:4regimes}c}.

At an even higher pump value, \(F_{\text{inc}} = 3.7\) (Fig.\hyperref[fig:4regimes]{~\ref*{fig:4regimes}d}), the system reaches a new stationary state, in which the central region exhibits a flat amplitude distribution with a regular phase, i.e. without singularities and sudden phase shifts. We therefore label this regime the \textit{superfluid regime}. In momentum space, a smooth distribution around \(\mathbf{k}=0\), is observed, with a
typical size determined by the distance between the two pumps.
\\

Before going further, we discuss the comparison with the usual Gross-Pitaevskii equation (GPE) description, which restricts the dynamics to the lower-polariton branch only. In our parameter range, the bare exciton and photon energies are resonant, resulting in polaritons that are, in the linear regime under resonant excitation at $k=0$, a $50{:}50$ mixture of exciton and photon.

As the exciton population increases, mean-field exciton-exciton interactions induce a blue shift of the excitonic resonance. This interaction-induced shift leads to an effective exciton-photon detuning, which modifies the exciton--photon composition of the polariton and progressively favors its photonic component. As a consequence, a regime is reached where the photonic density continues to increase with pump power, while the excitonic density exhibits a tendency to saturate \cite{Carusotto2013}. Such a redistribution between excitonic and photonic components cannot be captured within the standard lower-polariton Gross--Pitaevskii equation framework, where the polariton composition is fixed by construction. We quantify this effect by defining local photonic and excitonic fractions based on the corresponding relative densities as
\begin{equation}
    f_ {C,X}(\mathbf{r},t) =
    \frac{|\Psi_{C,X}(\mathbf{r},t)|^2}
         {|\Psi_C(\mathbf{r},t)|^2 + |\Psi_X(\mathbf{r},t)|^2},
\end{equation}

\begin{figure}[H]
    \centering
    \includegraphics[width=\linewidth]{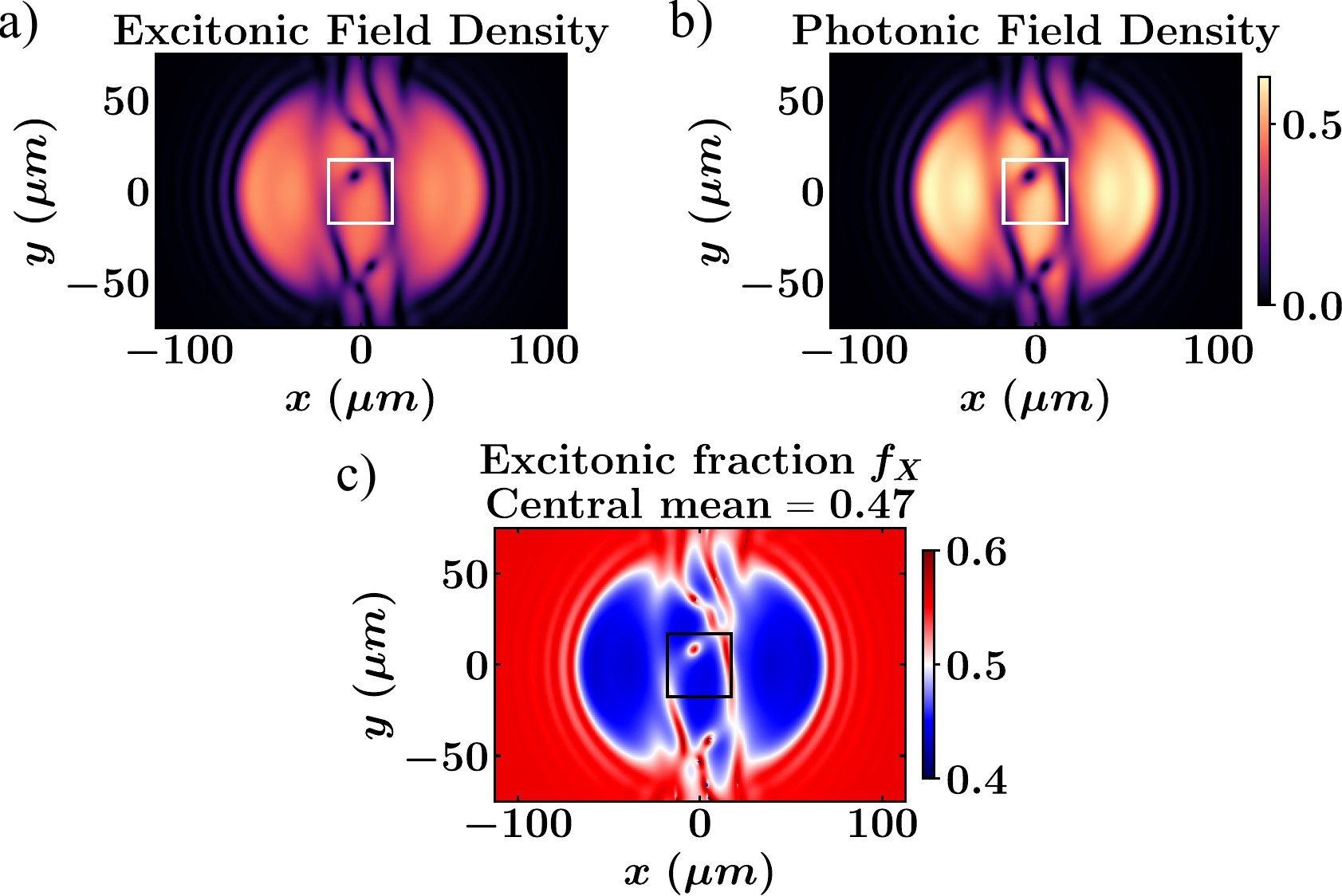}
    \caption{Comparison of excitonic and photonic field densities in our polariton system.
    Panels (a) and (b) show the spatial profiles of the excitonic and photonic field densities, respectively at a fixed time $t'=1980\,\tau_0'=1643\,\mathrm{ps}$.
    Both components exhibit the same spatial distribution, but the excitonic density is lower than the photonic one.
    Panel (c) shows the excitonic fraction $f_X$, which quantifies the matter--light composition of the quasiparticles.
    The central region (black box) highlights a balanced regime in which the excitonic fraction averages to $0.47$.}

    \label{fig:turbu_fraction}
\end{figure}

Figure~\hyperref[fig:turbu_fraction]{\ref*{fig:turbu_fraction}a,b} shows the excitonic and photonic densities at a fixed time for the same long-time dynamical regime as in Fig.~\hyperref[fig:4regimes]{\ref*{fig:4regimes}c}. From these densities, we extract the local excitonic fraction, displayed in Fig.~\hyperref[fig:turbu_fraction]{\ref*{fig:turbu_fraction}c}, which reveals pronounced spatial variations. Since this fraction directly controls the effective interaction strength in the system, its spatial and temporal dependence can significantly influence the resulting dynamics. This figure shows that polaritons are not exactly a $50{:}50$ exciton-photon mixture, and that a standard single-component Gross-Pitaevskii description would fail to provide a reliable quantitative predictions. The spatial and temporal dependence of the excitonic and photonic fractions therefore justifies the use of the coupled exciton-photon equations employed throughout this work.

Within the coupled equation description, we look for a quantitative criterion allowing us to distinguish the different regimes. To this end, we propose to analyze the competition between kinetic and interaction energies. 
We introduce the time-dependent, spatially averaged kinetic energy per polariton as
\begin{equation}
    \tilde E_{\rm kin}(t) =
    \frac{\displaystyle \int_S f_C(\mathbf{r},t) \,
    \bigl| \nabla \Psi_C(\mathbf{r},t) \bigr|^2 \, \text{d}\mathbf{r}}
    {\displaystyle \int_S |\Psi_C(\mathbf{r},t)|^2 \, \text{d}\mathbf{r}},
\end{equation}
with $S$ denoting the region of interest (yellow square in Fig.~\ref{fig:4regimes}).
Similarly, we evaluate the time-dependent and spatially averaged mean-field interaction energy per polariton as
\begin{equation}
    \tilde E_{\rm int}(t) =
    \left\langle
    f_X(\mathbf{r},t) \, |\Psi_X(\mathbf{r},t)|^2
    \right\rangle_{\mathbf{r}},
\end{equation}
Importantly, both energies are spatially averaged over the same region $S$, allowing for a consistent global comparison of the relative roles of kinetic and interaction effects across the different regimes.\\

Figure~\hyperref[fig:4regimes]{\ref*{fig:4regimes}} shows that, as the pump strength increases, the system progressively populates smaller momentum components, indicating a reduction of the kinetic energy. At the same time, the interaction energy, proportional to the exciton density \( |\Psi_X|^2 \), increases accordingly (not shown). This behavior reflects a competition between kinetic and interaction energies and can be quantified by defining the energy ratio \(\tilde \eta(t) = \tilde E_{\mathrm{int}(t)} / \tilde E_{\mathrm{kin}(t)}\).

\begin{figure}[t!]
    \centering
    \includegraphics[width=1\linewidth]{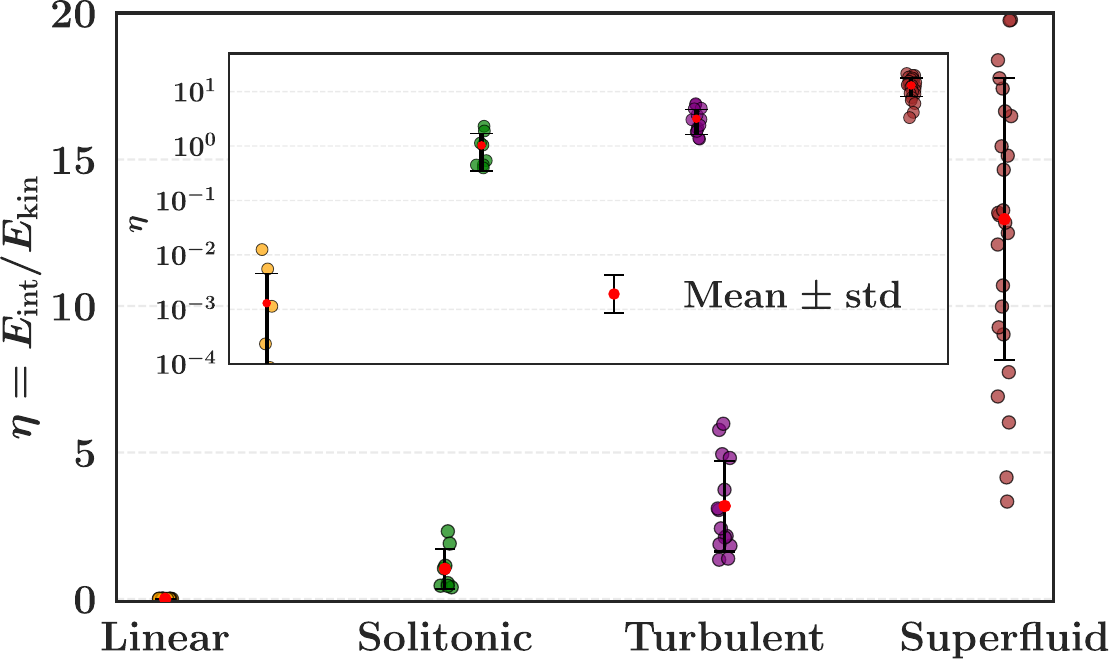}
    \caption{Linear-scale plot of the ratio between interaction and kinetic energy, accompanied by a panel showing the same ratio on a logarithmic ($y$-axis) scale. For each point, the ratio is evaluated over a time interval of $500\,\tau_0' \approx 400~\mathrm{ps}$, long compared to the dynamical timescale, allowing the turbulent dynamics to be properly averaged and removing time dependence. Red points indicate the mean value of the energy ratio for each regime, while black bars represent the standard deviation. Each point corresponds to a single simulation; the same simulations are used in Fig.\hyperref[fig:phasediagram]{~\ref*{fig:phasediagram}a}.
    }
    \label{fig:Histo}
\end{figure}

In the following, kinetic energies are time averaged over an interval of \(500\,\tau_0' \approx 400~\mathrm{ps}\), which is long compared to the characteristic dynamical timescale. This procedure allows the turbulent dynamics to be properly averaged and removes residual time dependence. In Fig.~\hyperref[fig:Histo]{\ref*{fig:Histo}d}, the ratio \(\eta = \langle\tilde  \eta(t) \rangle_t\) is plotted on both linear and logarithmic scales for clarity. We gather results from 70 different numerical simulations spanning the four regimes, with various \((\Delta, F_{\mathrm{inc}})\) parameters in the ranges \(\Delta \in [0,0.4] \leftrightarrow \Delta' \in [0,0.32]~\mathrm{meV}\) and \(F_{\mathrm{inc}} \in [0,3.7] \leftrightarrow I_{\mathrm{peak}} \in [0,2.29]~\mathrm{mW.\mu m^{-2}}\).\\

In the linear regime, interactions are negligible, and the injected energy is entirely stored in the kinetic contribution, corresponding to $\eta \ll 1$. In the solitonic regime, a reduction of the characteristic wave vector in momentum space is observed, indicating a redistribution of the injected energy toward interaction effects at the expense of the kinetic energy, as captured by the larger value $\eta = 1 \pm 0.7$. 
In the turbulent regime, the ratio is $\eta = 3.2 \pm 1.5$, and interactions are strong enough to destabilize solitonic states \cite{Koniakhin2019}, triggering the snake instability and leading to vortex nucleation. Although the interaction energy dominates over the kinetic energy in this regime, both remain of the same order of magnitude. 
At even higher densities, interaction effects strongly dominate: the system evolves toward a superfluid-like stationary state with $\eta\gg 1$, in which solitons are no longer supported. We note that the solitonic and turbulent regimes exist within the same range of energy ratio ($\eta = \mathcal{O}(1)$), so that they cannot be sharply distinguished using energetic criteria. Further details are provided in Appendix~\ref{app:Energy ratio}.\\

These regimes were previously reported in a configuration in which the polariton fluid encountered a localized obstacle~\cite{Simu_Pigeon}. In that work, vortex nucleation appeared in the low-density region behind the obstacle. In our geometry, two main differences emerge. First, vortex nucleation occurs within a high-density region, yielding a strong signal and providing a favorable setting to experimentally study turbulence. Second, unlike the flow case where a well-defined local Landau criterion can be applied, the counter-propagating configuration does not provide a well-defined fluid velocity in the central region. Consequently, no \textit{a priori} threshold can be formulated to predict whether superfluidity or a turbulent regime will occur. Nevertheless, as shown in Fig. \ref{fig:Histo}, we retain the idea that the competition between kinetic and interaction energies governs the transitions. An \textit{a posteriori} classification is therefore introduced to distinguish the regimes, in particular to separate stationary from turbulent dynamics.\\

\section{Phase diagram and turbulence mapping}

In this section, our aim is to identify quantitatively the region of existence of the turbulent regime in the system parameter space. We use the fact that turbulent regimes are characterized by nonstationary dynamics, whose magnitude can be quantified by introducing the time-averaged first-order coherence function \cite{Amo2011,Disorder_Amelio}
\begin{equation} 
g^{(1)}(\mathbf{r}) = \frac{|\braket{\Psi_C(\mathbf{r},t)}_t|}{\sqrt{\braket{|\Psi_C(\mathbf{r},t)|^{2}}}_t}= \frac{\left| \frac{1}{N} \sum_{n=1}^{N} \Psi_C(\mathbf{r},n) \right|}{\sqrt{\frac{1}{N} \sum_{n=1}^{N} |\Psi_C(\mathbf{r},n)|^2}}, 
\end{equation}
where \(\Psi_C(\mathbf{r},n)\) is the photonic field at a fixed position \(\mathbf{r}\) (near the center of the interference region) and at time index \(n\). The function \(g^{(1)}(\mathbf{r})\) ranges from 0, corresponding to a field whose phase fluctuates uniformly on the complex circle over time, to 1, indicating a perfectly stationary field. An example of a $g^{(1)}(\mathbf{r})$ space distribution is provided in Appendix~\ref{app:1st_order_coherence} for clarity in a specific turbulent case. By averaging this quantity over space in the region of interest (central region defined by yellow square in Fig.~\ref{fig:4regimes}), we obtain a scalar "order parameter" \( g^{(1)}=\langle g^{(1)}(\mathbf{r}) \rangle_{\mathbf{r}} \) that serves as a diagnostic for the degree of stationarity. 
In the linear and solitonic regimes, the field remains temporally coherent, yielding \(g^{(1)} \simeq 1\). Turbulent dynamics impede the establishment of a stationary regime, providing a  $g^{(1)}$ function significantly below 1. In the superfluid regime, the stationary is restored, with \(g^{(1)} \simeq 1\).\\

Turbulence emergence across different parameters is probed by examining the dependence of $g^{(1)}$ on the pump amplitude $F_{\text{inc}}$ and the laser detuning $\Delta$ from the lower-polariton energy, at fixed pump wave-vector  \(k_p' = 0.4\,l_0'^{-1}\approx 0.29\mu m ^{-1}\). For each couple $(F_{\text{inc}},\Delta)$, we simulate the long-time dynamics (up to $t'=2000\tau_0'$) and extract the value of the temporal coherence $g^{(1)}$ averaged over $\Delta t'=100\tau_0'$. Figure\hyperref[fig:phasediagram]{~\ref*{fig:phasediagram}a} shows a phase diagram where the color scale represents the spatially averaged coherence \(g^{(1)}\), with bright regions indicating reduced coherence. The simulations are only performed for the white points indicated in the diagram, and an interpolation is done in between the points for clarity. Using the classification introduced in Fig.\ref{fig:4regimes}, based on the spectrum, phase, and density profile, we label the regimes where the $g^{(1)}\approx1$ as either superfluid, solitonic, and linear, depending on the presence of fringes, few solitonic deeps or a flat profile in the density. Note that the four yellow points indicated on the line $\Delta=0.22$ correspond to the four cases associated with the four different regimes shown in Fig.~\ref{fig:4regimes}.

To investigate transitions between different regimes, we consider several paths across the diagram, marked by arrows. Green arrows correspond to a trajectory at fixed detuning $\Delta\approx0.22$, where the pump amplitude progressively increases. Along this path, the system passes through a reduced-coherence region between pump powers of \(F_{\text{inc}}\approx0.6\) and \(F_{\text{inc}}\approx2.4 \). Red arrows indicates a path at fixed pump power \(F_{\text{inc}}\approx1.5\), where the detuning increases. The turbulent region along this trajectory spans \(\Delta \approx0.07-0.36\).

At very low pump power, in the green area, the system resides in the linear regime. Here, the exciton density is too low, the interactions are negligible, and the interference pattern formed by the counter-propagating pumps remains stable over time. As the pump power increases, nonlinear effects become significant, giving rise to two solitonic phases (indicated in red) or to a turbulent regime. One solitonic phase appears at detuning above \(\Delta\approx0.22\), while the second occurs below \(\Delta\approx 0.09\).  Interestingly, no solitonic phase is observed for detunings between \(\Delta\approx 0.09\) and \(\Delta\approx 0.22\), which might be associated with stability constraints for solitons in driven-dissipative systems \cite{Smirnov2014}. Examination of stability criteria for solitons in our configuration with coupled-GPE description would allow to obtain a predictive power on the establishment of turbulent regimes, but goes beyond the aim of this paper. Finally, at pump powers above \(F_{\text{inc}}\approx3\) and detunings larger than \(\Delta\approx0.09\), interactions dominate, and the system reaches a stable superfluid regime.

\begin{widetext}
\centering

\begin{figure}[H]
\centering
\includegraphics[width=\textwidth]{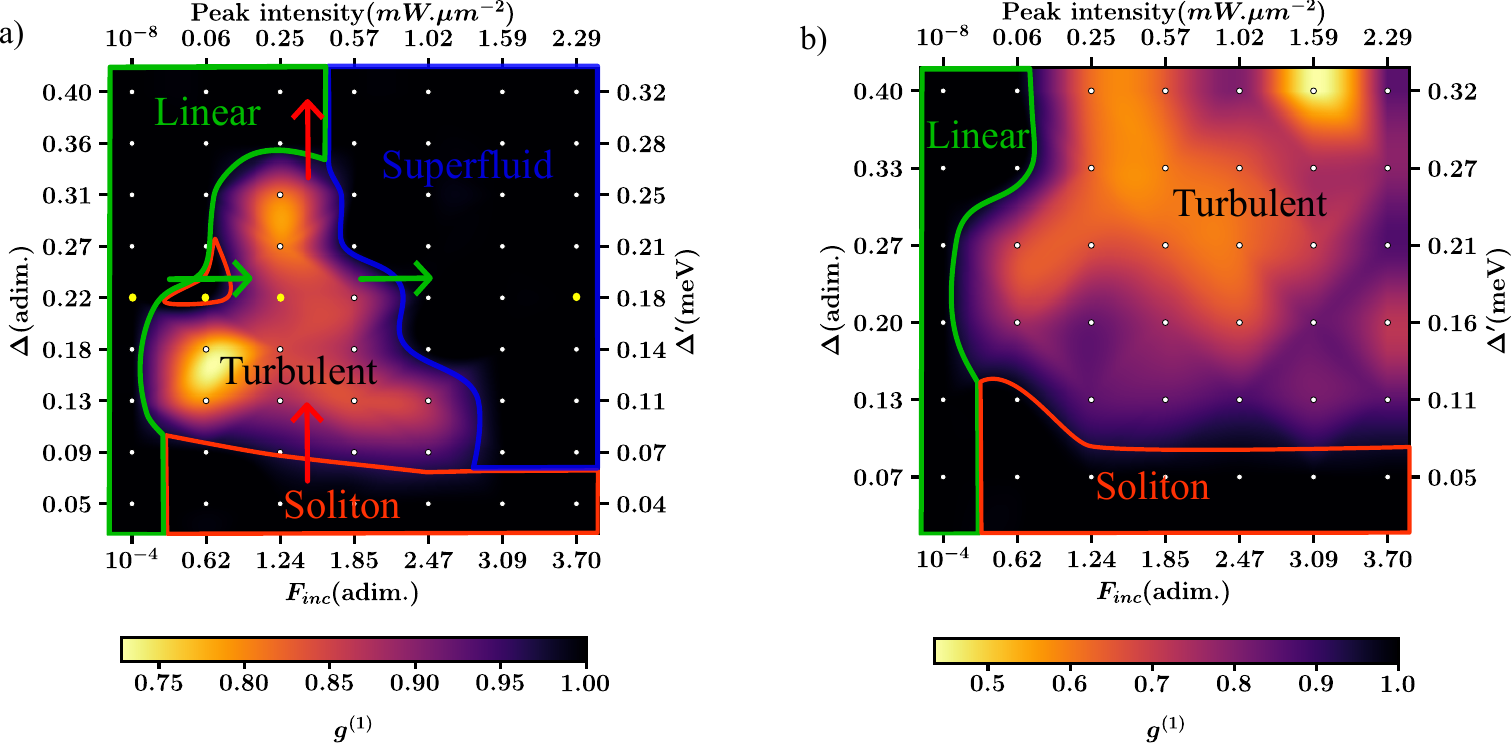}
\caption{Phase diagrams of the polariton fluid in the $(F_{\text{inc}},\Delta)$ plane.
(a) Phase diagram for the input wavevector $k_p' = 0.4\,l_0'^{-1} = 0.29\,\mu\mathrm{m}^{-1}$.
(b) Phase diagram for the input wavevector $k_p' = 0.6\,l_0'^{-1}= 0.43\,\mu\mathrm{m}^{-1}$. The color scale indicates the spatially averaged temporal coherence \(g^{(1)} \), which serves as a marker for stationarity and turbulence. Dark regions correspond to stationary regimes (linear, solitonic, superfluid), while bright zones indicate the emergence of turbulence. The zones are placed by hand to give an approximate indication of the different regimes. Each white points are a simulation and yellow points are these used for Fig.\ref{fig:4regimes}.
Parameters: Rabi splitting $2\hbar\Omega = 3.3$~meV, loss rates 
$\gamma = 0.02 \leftrightarrow \gamma' = 0.025~\mathrm{ps}^{-1}$, chosen photon wavelength $\lambda = 854~\mathrm{nm}$ and the interaction strength between excitons $\hbar g'_X = 0.003\,\mathrm{meV}\cdot\mu\mathrm{m}^2$.}
\label{fig:phasediagram}
\end{figure}
\end{widetext}

Transitions to the turbulent regime can occur in different ways. For some detunings (\(\Delta = 0.1 - 0.22\)), the system jumps directly from the linear to the turbulent regime. For others (\(\Delta = 0.22\), for example), the transition is gradual, passing through the solitonic regime. \\As shown by the energy analysis, turbulence appears when the interaction energy becomes comparable to, and slightly larger than, the kinetic energy, while remaining of the same order of magnitude. This condition can be reached by increasing the pump power beyond the solitonic regime, provided interactions are not yet saturated. Alternatively, the system may jump directly from the linear regime to a superfluid state, bypassing the solitonic regime altogether.\\

We also investigate a different scenario where the input pump momentum is larger and has the value: $k_p' = 0.6\,l_0'^{-1} \simeq 0.43\,\mu{\rm m}^{-1}$. This investigation is motivated by the fact that it is relevant experimentally, as $k_p$ offers an additional experimental controllable parameter. We thus perform similar numerical analysis (over roughly 40 different parameter configurations) and show the phase diagram for $k_p' = 0.6\,l_0'^{-1}$ in Fig.\hyperref[fig:phasediagram]{~\ref*{fig:phasediagram}b}, which is different compared to the $k_p' = 0.4\,l_0'^{-1}$ case of Fig.\hyperref[fig:phasediagram]{~\ref*{fig:phasediagram}a}. The turbulent region broadens markedly and extends toward both larger detunings $\Delta$ and higher pump amplitudes $F_{\text{inc}}$. In addition, the values of $g^{(1)}$ are substantially lower in this region, indicating a stronger and more developed turbulent dynamics. A stationary regime is still observed at low pump amplitudes and for detunings below $\Delta \simeq 0.09$, similarly to the lower-$k_p$ case. This quantitative change is mainly due to a different injection efficiency due to the input momentum of the pump associated with a different possibility for interaction to set in and counter-balance kinetic energy to establish a turbulent regime.

For completeness, we also examined the phase map for a smaller pump momentum $k_p' = 0.2\,l_0'^{-1} \simeq 0.14\,\mu{\rm m}^{-1}$. In this case, the turbulent region nearly disappears, reinforcing our interpretation: small $k_p$ does not prevent interactions, but it prevents the system from reaching the right "balance" between kinetic and interaction energy to support turbulence.\\

As a summary, the coherence maps in Fig.\hyperref[fig:phasediagram]{~\ref*{fig:phasediagram}a} reveal a clear re-entrant structure: the system can evolve from coherent to turbulent and back to coherent regimes (green and red arrows), following a \textbf{coherence–turbulence–coherence} sequence. The turbulent domain forms a distinct, connected region in the $(\Delta, F_{\text{inc}})$ parameter space, whose size and position are strongly controlled by the pump momentum. The pump separation also plays an important role. When the pumps are farther apart, achieving a balance between polariton density and kinetic energy becomes more difficult. Cavity losses and the effective density saturation induced by detuning and interactions limit the conditions under which turbulence can develop.

\section{Conclusion and Outlook}

In summary, we have numerically investigated the nonlinear dynamics of a two-dimensional exciton-polariton fluid under a counter-propagating coherent drive. By exploring the interplay between pump intensity and detuning, we identified a rich dynamical landscape featuring four distinct regimes: linear, solitonic, turbulent, and superfluid. These regimes arise from the balance between interaction and kinetic energies, and the coherence map provides a concise summary of the parameter regions (detuning and pump power) where turbulence is observed. Employing a realistic exciton--photon model that incorporates correlated disorder, we further identify robust signatures of these states across real-space, momentum-space, and temporal coherence observables.

It is important to note that in this work we have seen that solitons play an important roles to establish and maintain the turbulence regime. Indeed, the counter-propagating scheme enable the creation of instable solitons in the central region that continuously destabilize into vortex pair and emulate turbulent behavior. This process constitutes a natural higher-dimensional generalization of the phase-controlled bistability studied in 1D systems where  dark soliton trains were observed, such as those by Goblot \textit{et al.} \cite{Soliton_1D_counterProp}. While a longitudinal confinement smaller than the healing length (quasi-1D) can stabilize solitonic structures, the 2D geometry allows the existence of a parameter region where repeated fragmentations of solitons lead to the nucleation of 2D excitations (vortices) and thus the establishment of turbulent regime.

Regarding our modelization framework, we employ a minimalist exciton-photon coupled model that captures the essential physics without relying on arbitrary fitting parameters. This ensures that our results are transparent and directly comparable to experiments, with parameters—including photonic lifetimes and interaction strengths—fully compatible with state-of-the-art microcavity setups. Using this model, the coherence maps in Fig.\hyperref[fig:phasediagram]{~\ref*{fig:phasediagram}a} for an injection momentum $k_p' = 0.29~\mathrm{\mu m^{-1}}$ indicate the conditions under which turbulence arises. Specifically, turbulence occurs for detunings between $0.07$ and $0.28~\mathrm{meV}$ and for peak pump intensities from $0.01$ to $1.5~\mathrm{mW/\mu m^2}$, which corresponds to a total pump power between 35~mW and 2.6~W for the pump profile used in our simulations. These results demonstrate that turbulence should be experimentally accessible in this counter-propagating configuration over a wide range of parameters.

Taken together, our findings establish the counter-propagating geometry as a powerful and experimentally viable platform for studying out-of-equilibrium quantum fluids. By demonstrating that turbulence is a detectable and persistent phase within reachable parameter ranges, this work opens new avenues for systematic investigations of driven--dissipative turbulence. It also provides a strategy for exploring energy cascades, vortex clustering statistics, and universal scaling laws, bringing long-standing questions of quantum turbulence and energy transfer in mesoscopic light-matter fluids within reach of direct experimental observation.\\

\begin{acknowledgments}
We thank Iacopo Carusotto, Ivan Amelio, Jacqueline Bloch, Sylvain Ravets, and Nicolas Pernet for fruitful discussions. This work was supported by the European Research Council grant EmergenTopo (865151), the French government through the Programme Investissement d'Avenir (I-SITE ULNE /ANR-16-IDEX-0004 ULNE) managed by the Agence Nationale de la Recherche (ANR-25-CE57-2267-01), the Labex CEMPI (ANR-11-LABX-0007) and the region Hauts-de-France. This project has received funding from the European Union’s Horizon 2020 research and innovation programme under the Marie Skłodowska-Curie grant agreement No 101108433.
\end{acknowledgments}

\section*{DATA AVAILABILITY}  
The data that support the findings of this article are openly available \cite{DataAv}.

\bibliography{apssamp}

\appendix
\onecolumngrid     
\section{Coupled equations}

\label{app:adim}

The dynamics of the coupled cavity photon and exciton fields are described by the following driven-dissipative Gross--Pitaevskii equations:
\begin{equation}
\label{eq:CoupledDim_app}
\begin{aligned}
\rmi\hbar\frac{\partial\psi'_C}{\partial t'} 
&= \hbar\left(-\Delta'_C - \frac{\hbar \nabla'^{2}}{2m'_C} + W'(\mathbf{r})\right)\psi'_C 
- \frac{\rmi \hbar \gamma'_C}{2}\psi'_C 
+ \rmi\hbar\sqrt{\frac{\gamma'_C}{2}}\,F'_{\mathrm{inc}}(\mathbf{r}) 
- \hbar \Omega' \psi'_X,  \\\\
\rmi\hbar\frac{\partial\psi'_X}{\partial t'} 
&= \hbar\left(-\Delta'_X - \frac{\hbar \nabla'^{2}}{2m'_X} 
+ g'_X |\psi'_X|^2\right)\psi'_X 
- \frac{\rmi \hbar \gamma'_X}{2} \psi'_X 
- \hbar \Omega' \psi'_C .
\end{aligned}
\end{equation}
Here, $m'_C$ and $m'_X$ denote the effective masses of the cavity photon and exciton fields, respectively. The parameter $g'_X$ characterizes the strength of exciton--exciton interactions, while $\gamma'_C$ and $\gamma'_X$ represent the decay rates of photons and excitons. The detunings $\Delta'_{C,X} = \omega'_{C,X} - \omega'_p$ are defined with respect to the pump angular frequency $\omega'_p$, and $\Omega'$ denotes the Rabi splitting. The coherent pump field is described by $F'_{\mathrm{inc}}(\mathbf{r})$, which specifies the spatial profile of the external laser driving the cavity. The term $W'(\mathbf{r})$ represents a static disorder potential acting exclusively on the photonic field.

\subsection*{Dimensionless formulation}

To facilitate numerical simulations, we introduce dimensionless variables and parameters (denoted without a $'$, the counterparts with a $'$ have units) defined as
\begin{align*}
t &= \frac{\Omega'}{2}\, t', &
\mathbf{r} &= \sqrt{\frac{m'_C \Omega'}{\hbar}}\, \mathbf{r}', \\
\psi_{C,X} &= \sqrt{\frac{2g'_X}{\Omega'}}\,\psi'_{C,X}, &
F_{\mathrm{inc}} &= \frac{2\sqrt{g'_X}}{\Omega'}\, F'_{\mathrm{inc}}, \\
\gamma_{C,X} &= \frac{2}{\Omega'}\, \gamma'_{C,X}, &
\Delta_{C,X} &= \frac{2}{\Omega'}\, \Delta'_{C,X} .
\end{align*}

In addition, we define $\Delta' = E'_{p} - E'_{\mathrm{LP}}(k=0)$ as the energy detuning between the laser pump angular frequency and the lower polariton resonance at zero momentum. The characteristic timescale associated with Rabi oscillations is given by $\tau' = 2/\Omega'$, while the natural length scale of the system is the cavity length $l'_C = \sqrt{\frac{\hbar}{m'_C \Omega'}}$ and the rescaled disorder is $W(\mathbf{r})=\frac{2}{\Omega'}\, W'(\mathbf{r})$

Table~\ref{tab:Adim} summarizes the physical parameters and their corresponding dimensionless quantities, based on representative experimental values. In this work, we focus on the case where the bare exciton and photon energies are initially resonant, corresponding to the standard $50{:}50$ lower polariton composition. Injecting the rescaled variables of the Table 1 in Eq.\ref{eq:CoupledDim_app} allows obtaining the two coupled equations Eq.\ref{eq:CoupledEqADim_app}.

\begin{equation}
\label{eq:CoupledEqADim_app}
\begin{aligned}
\rmi\frac{\partial\psi_C}{\partial t} 
= &\left(-\Delta_C - \nabla^2 + W(\mathbf{r})\right)\psi_C 
- \rmi \frac{\gamma_C}{2} \psi_C 
+ \rmi \sqrt{\frac{\gamma_C}{2}} F_{\text{inc}}(\mathbf{r}) 
- 2\psi_X,\\
\rmi\frac{\partial\psi_X}{\partial t} 
=& \left(-\Delta_X + |\psi_X|^2\right)\psi_X 
- \rmi \frac{\gamma_X}{2} \psi_X 
- 2\psi_C.
\end{aligned}
\end{equation}

\begin{table}[H]
    \centering
    \renewcommand{\arraystretch}{1.4}
    \begin{tabular}{|c|c|}
        \hline
        \textbf{Rescaled variables} & \textbf{Parameters} \\
        \hline
        $t' = \frac{2}{\Omega_R} t = (0.80 \,\mathrm{ps})\,t$& $m'_C = 2.4\times10^{-5}m'_e$\\
        & $2\hbar\Omega' = 3.3\,\mathrm{meV}$\\
        $\mathbf{r'} = l'_C \mathbf{r} = (1.38\,\mu\mathrm{m})\,\mathbf{r}$&  \\
 $\nabla' = \nabla/l'_C = (0.72\,\mu\mathrm{m}^{-1})\,\nabla$&$\sigma'_W = 0.5\,\mu\text{m}$\\
        & $\hbar W'_0=1.3\,\mu\text{eV}$\\
        & $\hbar g'_X = 0.003\,\mathrm{meV}\cdot\mu\mathrm{m}^2$\\
 &\\
        $\psi'_C = \sqrt{\frac{\Omega'}{2g'_X}}\,\psi_C = (16.58\,\mu\mathrm{m}^{-1})\,\psi_C$& $\hbar\gamma'_C/2 = 8.2\,\mu\mathrm{eV}$\\
        $\psi'_X = \sqrt{\frac{\Omega}{2g'_X}}\,\psi_X = (16.58\,\mu\mathrm{m}^{-1})\,\psi_X$& $\hbar\gamma'_X/2 = 8.2\,\mu\mathrm{eV}$\\
        &$\hbar\gamma'_{\text{LP}}/2 = 8.2\,\mu\mathrm{eV}$\\
        & $\hbar \omega'_C = 1.65\,\mathrm{meV}$\\
        $|F'_{\text{inc}}|^2 = \left(\frac{\Omega'}{2\sqrt{g'_X}}\right)^2 |F_{\text{inc}}|^2$ & $\hbar\omega'_X = 1.65\,\mathrm{meV}$\\
        $= (345.15\,\mu\mathrm{m}^{-2}\cdot\mathrm{ps}^{-1})\,|F_{\text{inc}}|^2$& $\hbar\Delta'_{\text{LP}} = 0\,\mathrm{meV}$ \\
        \hline
    \end{tabular}
    \caption{Relations between dimensional and rescaled (dimensionless) physical quantities based on realistic experimental parameters.}
    \label{tab:Adim}
\end{table}

\section{Pump profile}
\label{app:pump_profile}
We consider that the spatial profile of the two pump beams can be given by:

\begin{figure}[H]
    \begin{minipage}{0.5\textwidth}
        \begin{equation}
        \begin{cases}
            P(\mathbf{r}) &= 
            \exp{\left(-\frac{x^2}{2\sigma_{x}^2} - \frac{y^2}{2\sigma_{y}^2}\right)}
            - 
            \exp{\left(-\frac{|x|^3}{2\sigma_{c}^3} - \frac{y^2}{2\sigma_{y}^2}\right)}, \\ \\
            F_{\text{inc}}(\mathbf{r}) &= F_{\text{inc}}
            P(x + d, y)\, e^{i \mathbf{k_p}\cdot\mathbf{r}} + F_{\text{inc}} P(x - d, y)\, e^{-i \mathbf{k_p}\cdot\mathbf{r}}
        \end{cases}
        \notag
        \end{equation}
    \end{minipage}
    \hfill
    \begin{minipage}{0.4\textwidth}
        \begin{minipage}{0.45\linewidth}
            \centering
            \includegraphics[width=\linewidth]{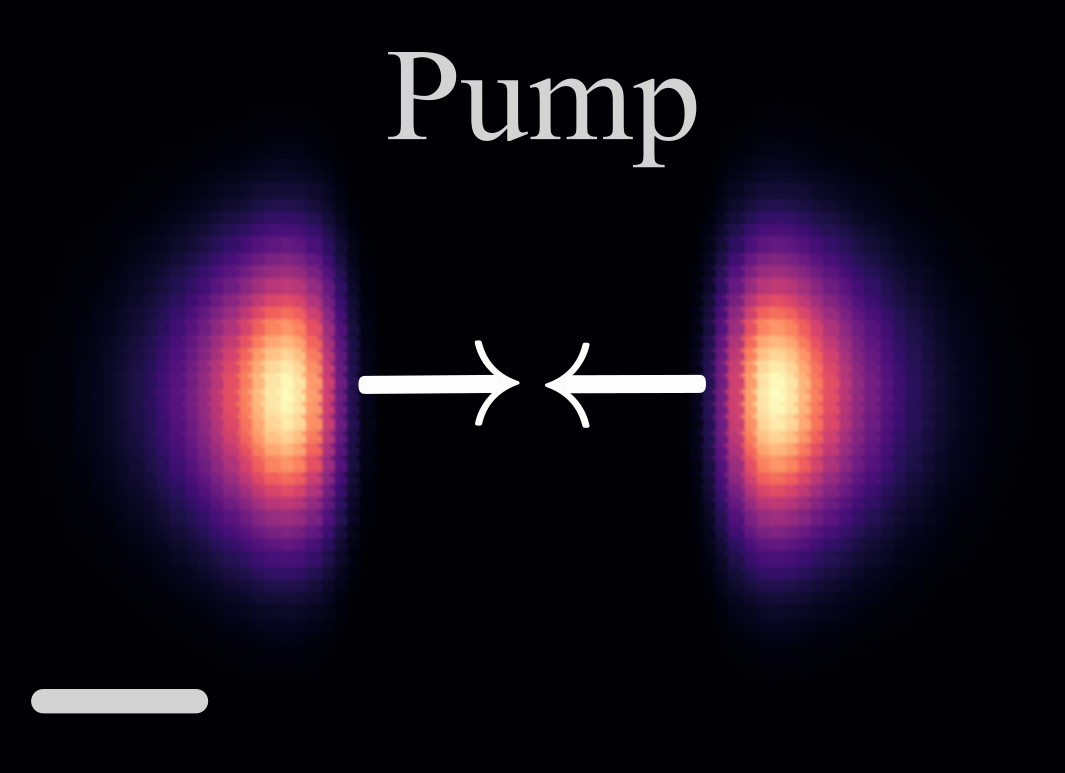}
        \end{minipage}
        \begin{minipage}{0.5\linewidth}
            \vspace{1em} 
            \caption{Pump profile. The scale bar is $\approx 33\,\mu\mathrm{m}$.}
            \label{fig:pump_only}
        \end{minipage}
    \end{minipage}
\end{figure}

Typical values used throughout this work are $\sigma_c'=40\,l_0' = 55~\mu\mathrm{m}$, $\sigma_x'=25\,l_0' = 35~\mu\mathrm{m}$, $\sigma_y' =20\,l_0'= 28~\mu\mathrm{m}$, $d' =7.5\,l_0'= 10~\mu\mathrm{m}$, and $k_p =0.4\,l_0'^{-1} = 0.29~\mu\mathrm{m}^{-1}$. Here, $\sigma_{x,y}'$ denote the widths of the uncut Gaussian pump spot along the $x$ and $y$ directions, while $\sigma_c'$ sets the characteristic length scale of the smooth cutting function. All parameters are expressed in dimensionless units in the numerical simulations; the values quoted above correspond to typical experimental orders of magnitude in GaAs-based microcavities.

The second term in $P(\mathbf{r})$ acts as a cutting function, enforcing a smooth yet rapid spatial decay over the length scale $\sigma_c$. Such steep pump profiles are commonly employed in polariton fluid dynamics to avoid imprinting the phase of the external laser onto the condensate and to allow the system to evolve according to its intrinsic nonlinear dynamics~\cite{Simu_Pigeon,pump_profile1}.

Finally, the two sides of the pump profile are separated by a distance $2d$, and opposite in-plane wavevectors $\pm\mathbf{k}_p$ are assigned to each side. Importantly, this type of configuration is readily achievable experimentally using spatial light modulators or tailored optical masks, which allow independent control of the cutting region, the pump separation $d$, and the injected momentum. We therefore consider a generic yet experimentally compatible pumping scheme, representative of current polariton-fluid experiments.

\section{Code robustness}
\label{app:code_verif}

To establish the reliability of the physical phenomena discussed in the main text, we conducted a comprehensive convergence analysis of the key numerical parameters. In particular, we verified that the observed dynamics are insensitive to variations in system size, spatial discretization, and temporal resolution, within the chosen numerical tolerances.

\begin{figure}[H]
     \centering
     \includegraphics[width=0.4\linewidth]{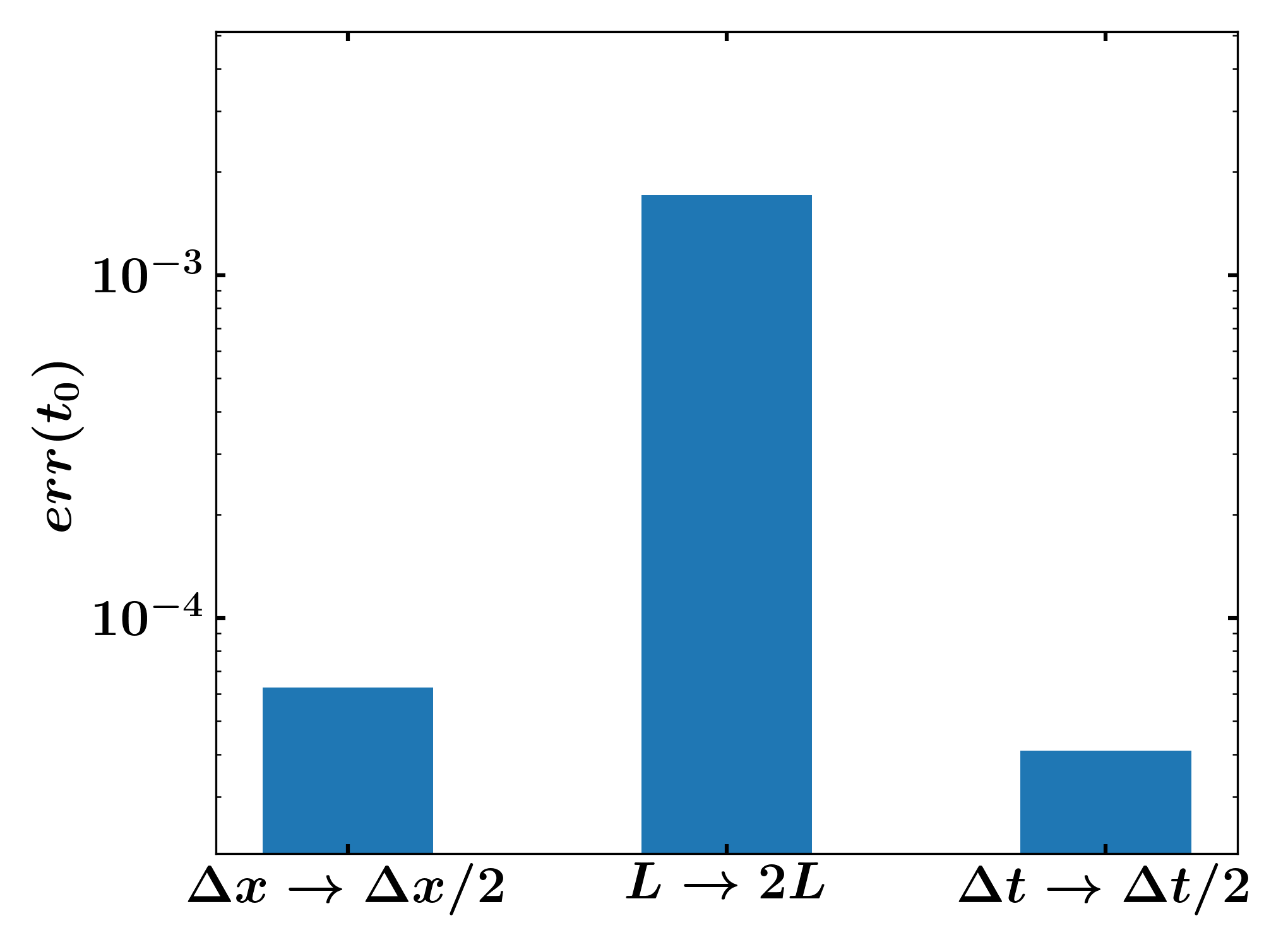}
     \caption{Sensitivity of the relative error at the fixed time
    \( t_0' = 2000\,\tau_0' = 1600\,\mathrm{ps} \) to the discretization parameters.
    The bars correspond to spatial refinement (\( \Delta x \rightarrow \Delta x/2 \)),
    increase of the system size at fixed spatial resolution (\( L \rightarrow 2L \)),
    and temporal refinement (\( \Delta t \rightarrow \Delta t/2 \)).
    }
     \label{fig:convergence}
 \end{figure}

Specifically, we independently varied the spatial step $\Delta x$, the time step $\Delta t$, and the system size $N$ (Fig.~\ref{fig:convergence}) to assess the robustness of the reported dynamical regimes. Relative errors were computed over a $10\times10$ central region, corresponding to the primary area of physical interest. The analysis focuses on the superfluid regime, which corresponds to the highest-density configuration and thus exhibits the strongest nonlinear effects. As a result, this regime provides the most stringent test of the numerical stability and accuracy of the simulations.

The relative error is defined as
\begin{equation}
err(t)
=
\frac{
\sum_{(i,j)\in S}
\left|\psi(t,i,j) - \psi_{\mathrm{ref}}(t,i,j)\right|^{2}
}{
\sum_{(i,j)\in S}
\left|\psi_{\mathrm{ref}}(t,i,j)\right|^{2}
},
\end{equation}
where $\psi_{\mathrm{ref}}$ denotes the reference solution, $t$ is a fixed time, and $S$ is the spatial region of size $N_x\times N_y$.

Reducing $\Delta x$ or $\Delta t$ by a factor of two produces deviations below $10^{-4}$, indicating that the results are well converged with respect to discretization. Increasing the system size $L$ by a factor of two yields a larger deviation on the order of $10^{-3}$, showing that finite-size effects are the dominant source of quantitative uncertainty. 

Importantly, in all cases considered, the system converges to the same dynamical regime with identical qualitative characteristics. This demonstrates that the observed phenomena are physically robust and not a consequence of numerical artifacts.

\section{Sneak instability}
\label{app:snake_instability}

\begin{figure}[H]
    \centering
    \includegraphics[width=0.9\linewidth]{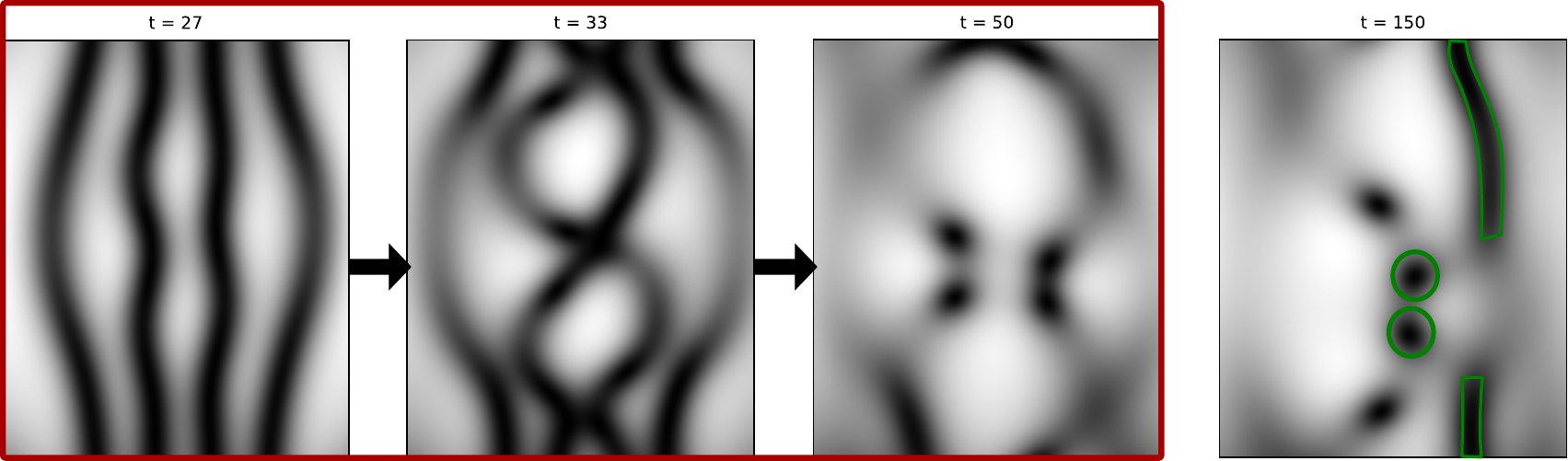}
    \caption{Example of the snake instability observed in the photonic density. The snapshots within the red rectangle show the first snake instability, where solitons destabilize and break into vortices. The last snapshot illustrates another instance of vortex generation through soliton destabilization: the soliton highlighted in green on the right breaks into two vortices, which are circled in green.
    Parameters: $1t=10\tau_0'$, $\Delta= 0.22$, $F_{\text{inc}}=1.2$ and $k_p=0.4$.
    }
    \label{fig:Snake_instability}
\end{figure}

The transition to a sustained turbulent regime is driven by a recurrent cycle of soliton formation and subsequent decay. Figure~\ref{fig:Snake_instability} illustrates the initial triggering of the snake instability ($t=27$ to $t=50$), where the transverse undulation of the dark soliton filaments leads to the nucleation of vortex–antivortex pairs.

Crucially, the dynamics do not stop after this first event. As shown in the final snapshot ($t=150$), the system continuously regenerates solitonic-like structures which, due to the persistent transverse instability, repeatedly undergo snaking and break down (highlighted in green). This cyclical process of soliton formation followed by destabilization acts as a sustained source of quantized vortices, thereby maintaining the system in a disordered, turbulent state. This feedback mechanism ensures that the dynamical features reported in the main text are not transient but correspond to a robust, self-sustaining regime.

\section{Energy ratio}
\label{app:Energy ratio}

\begin{figure}[H]
    \centering
    \includegraphics[width=0.8\linewidth]{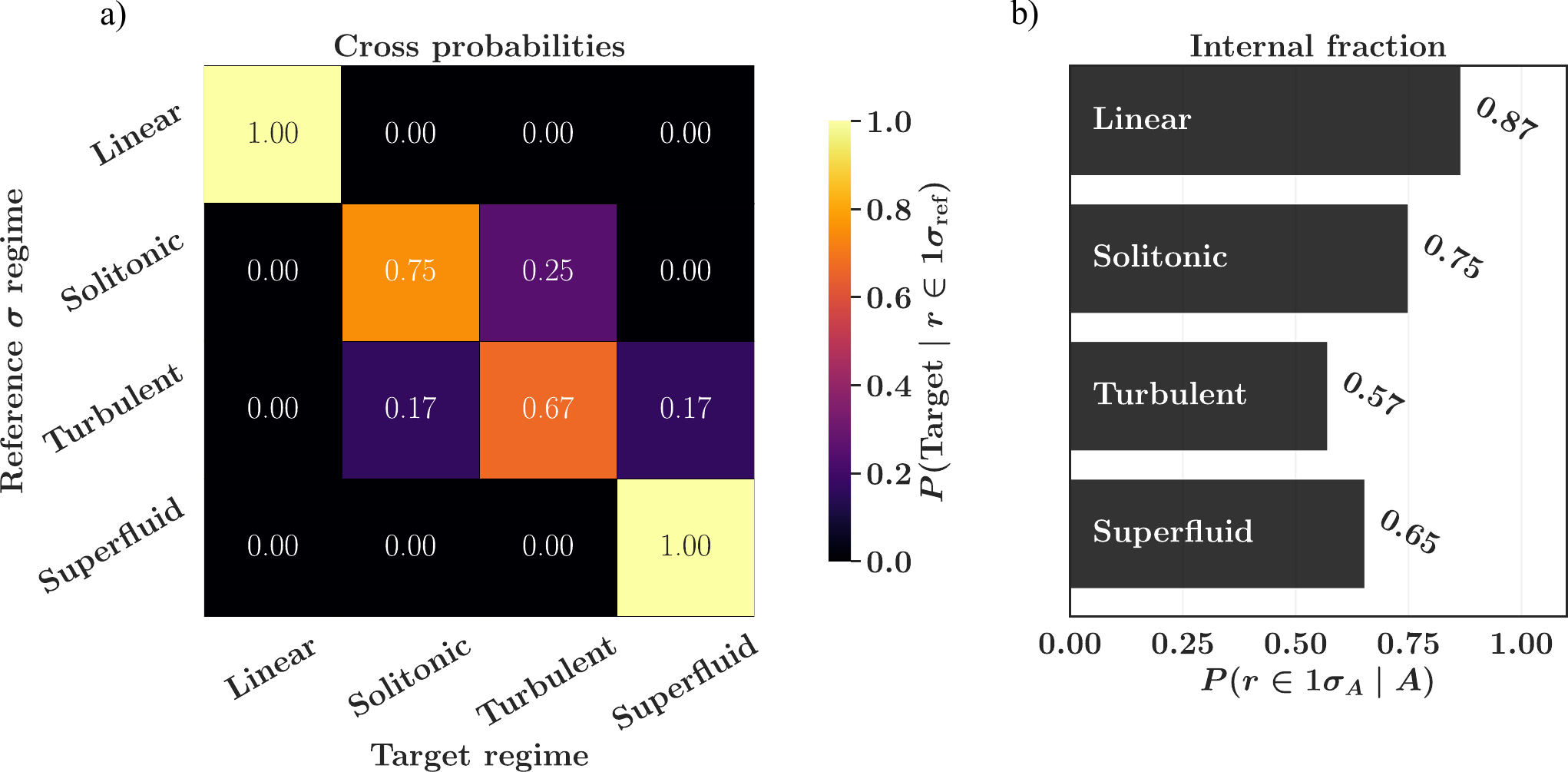}
    \caption{(a) Probability of identifying the target regime given that the energy ratio is selected within the range $\eta_{\mathrm{ref}} \pm \sigma_{\mathrm{ref}}$. 
    (b) Probability that the energy ratio lies within the interval $\eta_A \pm \sigma_A$, given that the selected ratio belongs to regime $A$.}
    \label{fig:Proba_energy}
\end{figure}

As discussed in Sec.~III, the energy ratio $\eta = E_{\mathrm{int}} / E_{\mathrm{kin}}$ provides an intuitive metric to characterize the different dynamical regimes. The linear and superfluid regimes correspond to well-defined limiting cases, where the dynamics is dominated by kinetic energy ($\eta \ll 1$) and interaction energy ($\eta \gg 1$), respectively. In contrast, the solitonic and turbulent regimes occupy an intermediate range of $\eta$ and exhibit a substantial overlap in their energy-ratio distributions.\\
This overlap is quantified in Fig.\hyperref[fig:Proba_energy]{~\ref*{fig:Proba_energy}a}, which displays the cross probabilities, defined as the probability of identifying a target regime when the energy ratio is chosen within the interval $\eta_{\mathrm{ref}} \pm \sigma_{\mathrm{ref}}$ of a reference regime. A key observation is that the linear and superfluid regimes are clearly separated from the others: the corresponding cross probabilities vanish for all non-reference regimes, confirming their role as distinct limiting cases. \\
By contrast, the solitonic and turbulent regimes are strongly mixed, indicating a significant ambiguity in their energetic classification. For instance, an energy ratio selected within the solitonic range is identified as turbulent in approximately $25\%$ of the cases, highlighting the strong overlap between these regimes.

Further insight is provided by Fig.\hyperref[fig:Proba_energy]{~\ref*{fig:Proba_energy}b}, which shows the fraction of realizations falling within the $\pm 1\sigma$ energy-ratio interval of each regime. This panel reveals that the dispersion of $\eta$ increases with its mean value, indicating increasingly large fluctuations of the energy ratio as interactions become more dominant.\\
Overall, Fig.~\ref{fig:Proba_energy} demonstrates that while the energy ratio $\eta$ is a powerful tool for a qualitative understanding of the different regimes, it does not allow for a sharp or unambiguous classification, particularly between the solitonic and turbulent regimes. This limitation motivates the use of the first-order coherence function as a more discriminating observable in the analysis.

\section{First-order coherence}
\label{app:1st_order_coherence}
To characterize the spatial coherence of the system in the turbulent regime, we compute the time-averaged first-order coherence function $g^{(1)}(\mathbf{r})$ over a selected region of interest. This analysis highlights how turbulence affects the local phase and density correlations, providing a clear signature of non-stationary dynamics.

\begin{figure}[H]
    \centering
    \includegraphics[width=0.5\linewidth]{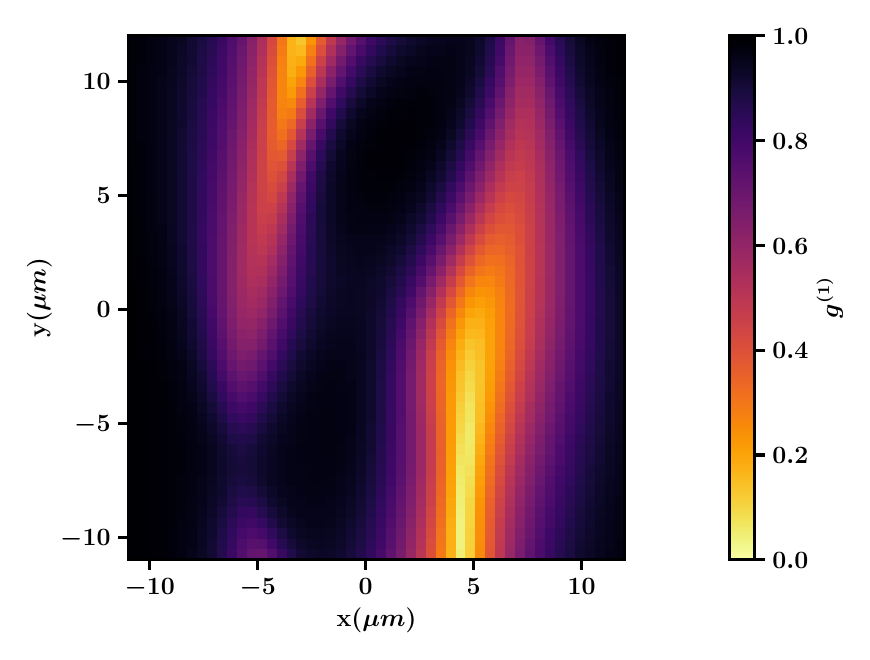}
    \caption{Spatial map of the first-order coherence function $g^{(1)}(\mathbf{r})$. The colormap represents the magnitude of $g^{(1)}(\mathbf{r})$ in the $(X,Y)$ plane.}
    \label{fig:G1_map}
\end{figure}

Figure~\ref{fig:G1_map} presents the time-averaged first-order coherence function $g^{(1)}(\mathbf{r})$, computed over $500\tau_0' \approx 400\,\mathrm{ps}$ in the turbulent regime and evaluated within the region of interest highlighted in yellow in Fig.~\ref{fig:4regimes}. Black areas correspond to stationary dynamics, whereas yellow areas reveal non-stationary behavior. In all other regimes, the region of interest used to compute the scalar coherence indicator is entirely black.

In the turbulent regime shown here, solitons, corresponding to reduced-density regions, dynamically break up into vortices. This process leads to a local reduction of the first-order coherence (yellow regions), with $g^{(1)}(\mathbf{r}) \approx 0$. In the remaining regions of space (black regions), the dynamics is mostly stationary during the integration time and resembles a superfluid regime, characterized by a nearly uniform density and a flat phase. As a result, the spatially averaged first-order coherence remains non-zero, with $\langle g^{(1)} \rangle \approx 0.75$.

\end{document}